\documentclass{iaus}
\usepackage{graphics}
  \checkfont{eurm10}
  \iffontfound
    \IfFileExists{upmath.sty}
      {\typeout{^^JFound AMS Euler Roman fonts on the system,
                   using the 'upmath' package.^^J}%
       \usepackage{upmath}}
      {\typeout{^^JFound AMS Euler Roman fonts on the system, but you
                   dont seem to have the}%
       \typeout{'upmath' package installed. iaus.cls can take advantage
                 of these fonts,^^Jif you use 'upmath' package.^^J}%
      }
  \else
  \fi

  \checkfont{msam10}
  \iffontfound
    \IfFileExists{amssymb.sty}
      {\typeout{^^JFound AMS Symbol fonts on the system, using the
                'amssymb' package.^^J}%
       \usepackage{amssymb}%

      }{}
  \fi

  \IfFileExists{amsbsy.sty}
    {\typeout{^^JFound the 'amsbsy' package on the system, using it.^^J}%
     \usepackage{amsbsy}}
    {}

\title[The Interplay among Black Holes, Stars and ISM in Galactic 
       Nuclei]{ Quasar Variability Measurements With SDSS
                    Repeated Imaging and POSS Data      }

\author[\v{Z}. Ivezi\'{c} {\it et al.\/}]%
{\v{Z}. Ivezi\'{c}$^{1,2}$
R.H. Lupton$^{1}$, M. Juri\'{c}$^{1}$, S. Anderson$^{2}$, 
P.B. Hall$^{1}$, G.T. Richards$^{1}$, C.M. Rockosi$^{2}$, 
D.E. Vanden Berk$^3$, E.L. Turner$^{1}$,
G.R. Knapp$^{1}$, J.E. Gunn$^{1}$, D. Schlegel$^{1}$, 
M.A. Strauss$^{1}$ and D.P. Schneider$^4$}

\affiliation{$^1$Princeton University, Princeton, USA, 
$^2$University of Washington, Seattle, USA,
$^3$University of Pittsburgh, Pittsburgh, USA,
$^4$The Pennsylvania State University, University Park, USA}

\pubyear{2004}
\volume{222}
\pagerange{1--8}
\date{?? and in revised form ??}
\jname{The Interplay among Black Holes, Stars and ISM \\in Galactic Nuclei}
\editors{Th. Storchi Bergmann, L.C. Ho \& H.R. Schmitt, eds.}
\begin{document}

\maketitle

\begin{abstract}
We analyze the properties of quasar variability using repeated 
SDSS imaging data in five UV-to-far red photometric bands, accurate to 
0.02 mag, for $\sim$13,000 spectroscopically confirmed quasars. 
The observed time lags span the range from 3 hours to over 3 years, 
and constrain the quasar variability for rest-frame time lags of 
up to two years, and at rest-frame wavelengths from 1000\AA\ to 
6000\AA. We demonstrate that $\sim$66,000 SDSS measurements of 
magnitude differences can be described within the measurement noise 
by a simple function of only three free parameters. The addition 
of POSS data constrains the long-term behavior of quasar variability
and provides evidence for a turn-over in the structure function.
This turn-over indicates that the characteristic time scale for 
optical variability of quasars is of the order 1 year.
\end{abstract}

{\bf \large \hskip -0.14 in 1.  Rest-frame Time Lag and Wavelength, and Luminosity Dependence}

Significant progress in the description of quasar variability has been recently made by 
employing SDSS data.  Vanden Berk et al. (2004, hereafter VB) used a combination of imaging 
and spectrophotometric magnitudes to investigate the correlations of 
variability with rest frame time lag (up to 2 years), luminosity, rest wavelength, 
redshift, the presence of radio and X-ray emission, and the presence of broad absorption 
line outflows. Variability on longer time scales was studied by
de Vries, Becker \& White (2003, hereafter dVBW) who compared SDSS and POSS measurements. 
We extend these studies by analyzing repeated SDSS imaging data. Here we summarize the main 
results obtained for a sample of $\sim$13,000 quasars 
with $\sim$66,000 magnitude difference measurements in 5 SDSS bands\footnote{We do not employ
spectrophotometric magnitudes as they are 3-4 times less accurate than imaging magnitudes, 
and are not available for $u$ and $z$ bands.} (for a detailed description 
of SDSS survey see Stoughton et al. 2001, and references therein), and recalibrated 
POSS I and POSS II photometry (Sesar et al. 2004). Details and supporting analysis will be 
presented elsewhere (Ivezi\'{c} et al. 2004, in prep).

We find that SDSS magnitude difference measurements, $\Delta m$, follow an exponential 
distribution, $p(\Delta m)\propto exp(-|\Delta m|/\Delta_c)$, where the characteristic
variability scale, $\Delta_c$ is a function of rest-frame time lag ($\Delta t_{RF}$, days), 
wavelength ($\lambda_{RF}$, \AA), and absolute magnitude in $i$ band ($M_i$). The 
variability scale is related to the more commonly used structure function (see dVBW
for definition) by $SF=\sqrt{2}\Delta_c$; we find that the function

\vskip  0.03in
\hskip -0.14in 
 $\phantom{mala} SF = (1.00\pm0.03)\,[1 + (0.024\pm0.04)\,M_i] \, (\Delta t_{RF} / \lambda_{RF})^{0.30\pm0.05} 
\phantom{mirelajezlocesta} (1.1)$
\vskip  0.03in
\hskip -0.14in
describes $\Delta m$ measurements to within the measurement noise ($\sim$0.02 mag).
Note that there is no dependence on redshift (see Figure 1). The difference between
the data and this model (right panel)
 reveals a feature at $\sim$2800 \AA, which is coincident with 
the MgII line. This anti-correlation between the continuum and Mg II line variability is 
indeed known from reverberation mapping -- due to the excellent SDSS photometry, we 
recovered this result using broad-band data!

\begin{figure}
\centering
\resizebox{18cm}{!}{\includegraphics{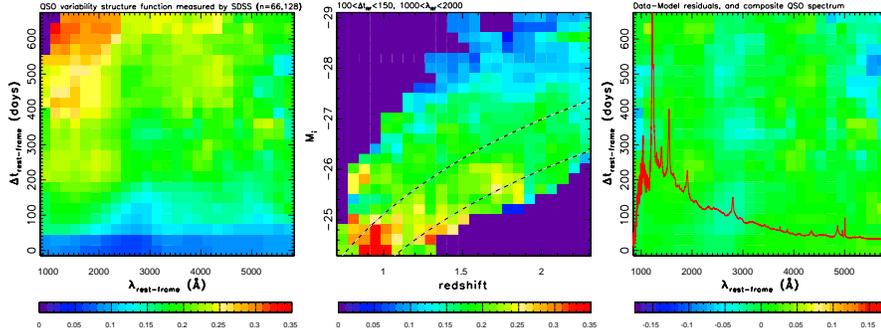}} 
\caption{The left panel displays the measured structure function as a function of rest-frame 
time lag and wavelength (each pixel contains $\sim25$ objects). The middle panel
shows the structure function in a narrow range of rest-frame time lag and wavelength
(100 d$<\Delta t_{RF}<$150 d, 1000\AA $<\lambda_{RF}<$2000 \AA),
as a function of redshift and luminosity. The lines of constant variability are nearly 
parallel to the redshift axis, suggesting that the dependence of variability
on luminosity is much stronger than the dependence on redshift. The right panel
displays the difference between the data shown in the left panel and the best-fit
model described by eq.(1.1). The red line shows the composite quasar spectrum 
from Vanden Berk et al. (2001). The negative residuals
at $\sim$2800 \AA\ are due to the anti-correlation of continuum and MgII line
variability.
}
\label{fig}
\end{figure}

{\bf \large \vskip 0.1in \hskip -0.14 in 2.  Detection of a Turn-over in the Structure Function}

\begin{figure}
\centering
\resizebox{10cm}{!}{\includegraphics{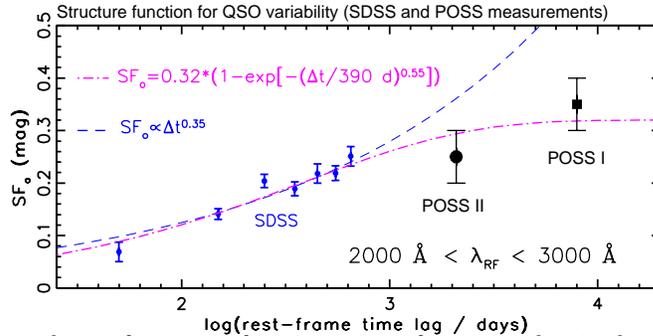}} 
\caption{The dependence of structure function on rest-frame time lag, 
in the wavelength range 2000--3000 \AA, for two data sets: SDSS-SDSS for 
short time lags (small symbols), and SDSS-POSS for long time lags (large symbols).
}
\label{fig}
\end{figure}

The time lags between Palomar Observatory Sky Surveys (POSS I and POSS II) and SDSS  
are much longer (up to $\sim$50 years), than spanned by the available SDSS repeated 
imaging data, and thus offer a possibility of detecting deviations from the simple 
power-law measured for short time scales using repeated SDSS imaging.
The observed SDSS-POSS long-term variability is {\it smaller} than predicted by the 
extrapolation of the power-law from eq.~1.1 (Sesar et al. 2004). The observations 
(Fig. 2) are well described~by 

\vskip  0.03in
\hskip -0.14in 
 $\phantom{mala}  SF(\Delta t_{RF}) = D\, \left(1 - {\rm e}^{-(\Delta t_{RF} / \tau)^\gamma}\right) 
\phantom{mirelajejakojakopunopunojakozlocestaX} (2.1)$
\vskip  0.03in
\hskip -0.14in
with $D=0.32\pm0.03$, $\tau=(390\pm80)$ days, and $\gamma=0.55\pm0.05$. This best-fit is 
shown in Figure 2 by the dot-dashed line. We conclude that  {\it the characteristic time 
scale for optical variability of quasars is of the order 1 year.}


\vskip -0.25in
\begin{thebibliography}{}
\vskip -0.1in
\bibitem[]{} de Vries, W.H., Becker, R.H., \& White, R.L. 2003, AJ, {\bf 126}, 1217
\bibitem[]{} Sesar, B., Svilkovi\'{c}, D., Ivezi\'c, \v Z., and 14 co-authors 2004, astro-ph/0403319
\bibitem[]{} Vanden Berk, D.E., Wilhite, B.C., Kron, R.G., and 11 co-authors 2004, ApJ, {\bf 601}, 692
\bibitem[]{} Vanden Berk, D.E., Richards, G.T., Bauer, A., and 59 co-authors 2001, AJ, {\bf 122}, 549 
\bibitem[]{} Stoughton, C., Lupton, R.H., Bernardi, M., and 189 co-authors 2002, AJ, {\bf 123}, 485
\end{thebibliography}
\end{document}